\pdfoutput=1 
\documentclass[doublecol,a4paper]{epl2} 

\usepackage[english,francais]{babel}

\usepackage{amsfonts,amsmath,amssymb,ifpdf,epsfig,graphicx}
\usepackage{amsmath,amssymb,mathrsfs}

\renewcommand{\leq}{\leqslant}
\renewcommand{\geq}{\geqslant}

\def\eqlaw{\stackrel{\mbox{\tiny (law)}}{=}}     
\newcommand{\ket}[1]{|\kern.3ex#1\kern.3ex\rangle}
\newcommand{\bra}[1]{\langle\kern.3ex #1 \kern.3ex|}

\newcommand{\mean}[1]{\left\langle #1\right\rangle}

\newcommand{\EXP}[1]{e^{#1}}         
\newcommand{\argcosh}{\mathop{\mathrm{argcosh}}\nolimits}

\newcommand{\im}{\mathop{\mathrm{Im}}\nolimits}      
\newcommand{\tr}[1]{\mathop{\mathrm{tr}}\nolimits\left\{ #1 \right\}}  
\newcommand{\sign}{\mathop{\mathrm{sign}}\nolimits}  

\def\I{{\rm i}}

\newcommand{\derivp}[2]{\frac{\partial #1}{\partial #2}}

\def\D{{\rm d}}                  

\def\Nc{N}

\def\HDirac{\mathcal{H}}
\def\conduc{\mathcal{G}}

\newcommand{\abs}[1]{\ensuremath{\left| #1 \right|}}
\newcommand{\pf}{\ensuremath{\mathrm{pf}}}

\begin{document}

\selectlanguage{english}

\title{Topological phase transitions in the 1D multichannel Dirac equation with random mass and a random matrix model}

\author{Aur\'elien Grabsch\inst{1,2,3}\thanks{\email{aurelien.grabsch@u-psud.fr}} \and 
Christophe Texier\inst{1}\thanks{\email{christophe.texier@u-psud.fr}}}
\shortauthor{A. Grabsch and C. Texier}

\institute{                    
  \inst{1} LPTMS, CNRS, Univ. Paris-Sud, Universit\'e Paris-Saclay, 91405 Orsay cedex, France.
  \\
  \inst{2} \'{E}cole Normale Sup\'{e}rieure de Cachan, 
 94235 Cachan cedex, France.
  \\
  \inst{3} School of Mathematics, University of Bristol, Bristol BS8 1TW, United Kingdom.
  }







\abstract{ 
We establish the connection between a multichannel disordered model --the 1D Dirac equation with $N\times N$ matricial random mass-- and a random matrix model corresponding to a deformation of the Laguerre ensemble.
This allows us to derive exact determinantal representations for the density of states and identify its low energy ($\varepsilon\to0$) behaviour $\rho(\varepsilon)\sim|\varepsilon|^{\alpha-1}$.
The vanishing of the exponent $\alpha$ for $N$ specific values of the averaged mass over disorder ratio corresponds to $N$ phase transitions of topological nature characterised by the change of a quantum number (Witten index) which is deduced straightforwardly in the matrix model.
}

\maketitle


Since the pioneering work of Dorokhov-Mello-Pereyra-Kumar (DMPK) \cite{Bee97,MelKum04}, models of multichannel disordered wires have played a prominent role in the theory of disordered systems as they allow to describe a situation intermediate between the strictly one-dimensional (1D) case and higher dimensions. 
With the dimensionality, another crucial aspect of disordered systems is the presence or not of symmetries, what may strongly affect the localisation properties. This has led to the classification within orthogonal, unitary and symplectic classes, depending on the existence of time reversal and spin rotational symmetries, according to the Wigner-Dyson classification for random matrices, labelled with the Dyson index $\beta\in\{1,\,2,\,4\}$.
These three classes were later completed by three classes with an additional chiral symmetry and four Bogoliubov-de Gennes (BdG) classes for disordered superconductors \cite{Zir96,AltZir97} (see also \cite{EveMir08}).
Several models of multichannel wires in these symmetry classes were studied by Brouwer and coworkers by extending the DMPK approach \cite{BroMudSimAlt98,BroMudFur00,BroMudFur01,TitBroFurMud01}.
The interest for such models has been recently renewed, as they can support topologically protected Majorana zero modes \cite{Kit01,OreRefOpp10}, 
which are stable against perturbations, like a small amount of disorder, what could be used for quantum computation~\cite{Kit01,AliOreRefOppFis11}. 
When sufficiently strong, disorder can however drive topological phase transitions (quantum phase transitions associated with the change of a quantum number of topological nature)~\cite{MotDamHus01,PotLee10,RieBroAda13}. 

Although multichannel models can be studied from a symmetry viewpoint 
\cite{LudSchSto13,MorFurMud15}, detailed analysis of specific models are also useful, in particular for the determination of nonuniversal properties (microscopic parameter dependences)~\cite{RieBro14}.
In this letter, we propose such a detailed study of 1D models belonging to the chiral classes of disordered systems, and establish the connection with a random matrix model defined by the following matrix distribution
\begin{align}
  \label{eq:StatDistfZ}
  f(Z) =& \, \mathcal{C}_{\Nc,\beta}^{-1}\, \left(\det Z \right)^{-\mu-1-\beta(\Nc-1)/2}
  \nonumber\\
  &\times\exp\left[
    -\frac12 \tr{ G^{-1}(Z +k^2 Z^{-1}) }
  \right]
\end{align}
with matrix argument in the group of $\Nc\times\Nc$ Hermitian matrices with real ($\beta=1$), complex ($\beta=2$) or quaternionic ($\beta=4$) elements and positive eigenvalues. 
Integration over this group will be denoted $\int_{Z>0}\mathrm{D}Z\,f(Z)=1$, thus $\mathcal{C}_{\Nc,\beta}$ ensures normalisation.
We will show that several interesting properties of the 1D Dirac equation with random mass (spectral properties and topological index) can be straightforwardly obtained from this matrix distribution.
In the ``isotropic case'', when $G=g\mathbf{1}_\Nc$, the distribution is \textit{invariant} under unitary transformations. Then, Eq.~\eqref{eq:StatDistfZ} interpolates between the Laguerre distribution $L_{\Nc,\theta}(Z)\propto(\det Z)^{\theta-1}\,\exp\big[-\tr{Z}\big]$ with $\theta>0$ (obtained when $k\to0$) and the ``inverse-Laguerre'' distribution $I_{\Nc,\theta}(Z)\propto(\det Z)^{-\theta-\beta(\Nc-1)-1}\,\exp\big[-\tr{Z^{-1}}\big]$ (obtained when $k\to\infty$).
The breaking of the invariance of $f(Z)$ under unitary transformations, when $G$ is not the identity matrix, is a further deformation of the distribution. 
Whereas most analytical studies of multichannel disordered models take as a crucial asumption the isotropy among the channels, some of our results will not rely on this hypothesis.

The outline of the letter is as follows~:
we define our model and introduce the scattering problem. Then we explain how the distribution \eqref{eq:StatDistfZ} appears.
This will provide the ground from which localisation and spectral properties will be recovered directly (in the isotropic case).
The topological phase transitions, identified from the spectral properties, will be characterised through the calculation of a topological quantum number, achieved directly from a detailed analysis of the matrix distribution~\eqref{eq:StatDistfZ}.


\section{The model}

The disordered model is the 1D Dirac equation $\HDirac\Psi(x)=\varepsilon\Psi(x)$ for $2\Nc$ component spinors with
\begin{equation}
  \label{eq:HDirac}
  \HDirac = \I \, \sigma_2\otimes \mathbf{1}_\Nc \, \partial_x + \sigma_1\otimes M(x)
  \:,
\end{equation}
$\sigma_i$ being a Pauli matrix. 
The Hamiltonian exhibits a chiral symmetry $\sigma_3\HDirac\sigma_3=-\HDirac$, which puts the problem in one of the three chiral classes, depending whether the random $\Nc\times\Nc$ Hermitian matrix $M(x)$ has real ($\beta=1$), complex ($\beta=2$) or quaternionic ($\beta=4$) elements. 
The mass is uncorrelated in space, distributed according to
\begin{equation}
  P[ M(x) ] \propto 
  \EXP{
    - (1/2)\int\D x \, \mathrm{tr}\{ (M(x)-\mu\,G)^\dagger G^{-1}(M(x)-\mu\,G) \}
  }
  \:.
\end{equation} 
The correlations and the mean value $\mean{M(x)}=\mu\,G$ are controlled by the \textit{same} real symmetric matrix $G$. 
The dimensionless parameter $\mu$ is the averaged mass over disorder ratio and will play a central role as it drives the topological phase transitions. 


\section{Scattering problem, Riccati matrix and random matricial process}

Our starting point is a scattering formulation~:
we consider the Dirac equation on the half line with mass $M(x)$ vanishing for $x>L$ in order to set a scattering problem. 
We find convenient to gather the $\Nc$ independent solutions of the Dirac equation in a $2\Nc\times\Nc$ ``spinor'', $\Psi=(\varphi^\mathrm{T},\chi^\mathrm{T})^\mathrm{T}$ where the two ``components'' $\varphi$ and $\chi$ are two $\Nc\times\Nc$ matrices.
In the free region ($x>L$), we can write the spinor as the superposition of incoming and outgoing plane waves.
For $\varepsilon>0$ we have~:
\begin{equation}
  \label{eq:SSS}
  \Psi(x) = 
  \begin{pmatrix}
    \phantom{\I}\,\mathbf{1}_\Nc \\ \I\,\mathbf{1}_\Nc
  \end{pmatrix}\EXP{-\I\varepsilon(x-L)}
  +
  \mathcal{S}(\varepsilon)\otimes\mathbf{1}_2
  \begin{pmatrix}
    -\mathbf{1}_\Nc \\ \I\,\mathbf{1}_\Nc
  \end{pmatrix}\EXP{+\I\varepsilon(x-L)}
  \:,
\end{equation}
where $\mathcal{S}(\varepsilon)$ is the $\Nc\times\Nc$ scattering matrix (here characterising the total reflection).
Because the chiral symmetry relates positive and negative energies, we will always choose $\varepsilon>0$ (see Ref.~\cite{Bee15} for a discussion of the $\mathcal{S}$-matrix symmetry). 
The study of the strictly 1D case ($\Nc=1$) has emphasized the role of a Riccati variable for providing the spectral and localisation informations \cite{BouComGeoLeD90,ComTexTou11,ComLucTexTou13,GraTexTou14}. 
We extend here this analysis to the multichannel case and emphasize the connection with the scattering problem.
We define the Riccati matrix as $Z_\varepsilon=-\varepsilon\,\chi\,\varphi^{-1}$, which obeys the stochastic differential equation (SDE)~:
\begin{equation}
  \label{eq:RiccatiEq}
  \partial_x Z_\varepsilon(x) = -\varepsilon^2 - Z_\varepsilon(x)^2 - M(x) Z_\varepsilon(x) - Z_\varepsilon(x) M(x) 
  \:.
\end{equation} 
The initial condition for $\Psi(x)$ is chosen such that the chiral symmetry is preserved~:
we will choose either $\varphi(0)=0$ (which corresponds to $M(x)\to-\infty$ on $\mathbb{R}_-$) or $\chi(0)=0$ ($M(x)\to+\infty$ on $\mathbb{R}_-$).~\footnote{
  A more general boundary condition ensuring the confinement of the particle on $\mathbb{R}_+$ is
  $\big(\EXP{2\I\theta\sigma_2}+\sigma_3\big)\Psi(0)=0$  \cite{AntComKne90,TexHag10} i.e.
  $\cos\theta\,\varphi(0)+\sin\theta\,\chi(0)=0$. 
  Only $\theta=0$ or $\pi/2$ preserves the chiral symmetry.} 
In this case, the Riccati matrix is Hermitian.
Matching of \eqref{eq:SSS} at the boundary reads
$\chi(L)\,\varphi(L)^{-1}=\I(\mathbf{1}_\Nc+\mathcal{S})(\mathbf{1}_\Nc-\mathcal{S})^{-1}$ and allows to express the scattering matrix in terms of the Riccati matrix~:
\begin{equation}
  \label{eq:MatrixR}
  \mathcal{S}(\varepsilon) = 
  \left[ \varepsilon - \I\, Z_\varepsilon(L) \right]\left[ \varepsilon + \I\, Z_\varepsilon(L) \right]^{-1}
  \:.
\end{equation}

The SDE \eqref{eq:RiccatiEq} may then be related to a Fokker-Planck equation (FPE) describing the matricial random process. 
Without any further assumption (like the isotropy assumption assumed in \cite{Bee97,BroMudSimAlt98,BroMudFur00,BroMudFur01,TitBroFurMud01,LudSchSto13}), setting a purely imaginary energy $\varepsilon=\I k\in\I\mathbb{R}$, we have obtained (only for $\beta=1$ and $2$) that the stationary distribution for the stochastic matricial process is given by Eq.~\eqref{eq:StatDistfZ} (cf.~\cite{SM}).
When $\Nc=1$, we check that (\ref{eq:StatDistfZ},\ref{eq:RiccatiEq}) correspond to the known results \cite{BouComGeoLeD90,ComTexTou11}.

We can make more explicit the relation with the distribution of the analytically continued scattering matrix $\mathcal{S}(\I k)=(k-Z_{\I k})(k+Z_{\I k})^{-1}$ with eigenvalues in the interval $[-1,+1]$.
The Jacobian of the transformation is 
$\mathrm{D}Z=\mathrm{D}\mathcal{S}\,(2k)^{\Nc(1+\beta(\Nc-1)/2)}\,\det(1-\mathcal{S})^{-2-\beta(\Nc-1)}$, 
thus this shows that the scattering matrix distribution is a deformation of the Jacobi distribution~:
$
P(\mathcal{S})\propto
\det(1+\mathcal{S})^{-\mu-1-\beta(\Nc-1)/2}
\det(1-\mathcal{S})^{\mu-1-\beta(\Nc-1)/2}
\exp\big[-k\tr{G^{-1}(1+\mathcal{S}^2)(1-\mathcal{S}^2)^{-1}}\big]
$.


\section{Lyapunov spectrum at \mathversion{bold}$\varepsilon=0$\mathversion{normal}}

It is well-knwon that the introduction of any small amount of disorder, uncorrelated in space, leads to the complete localisation of the eigenstates. 
For weak disorder, $|\varepsilon|\gg g$, the localisation properties of the model \eqref{eq:HDirac} are expected to fall into the standard universality classes studied in \cite{Bee97}. 
This does not prevent delocalisation to occur at specific points of the spectrum as a consequence of some symmetry.
The Dirac equation may indeed present a delocalisation point, exactly at $\varepsilon=0$, as a consequence of the chiral symmetry. The energy can therefore be viewed as the chiral-symmetry breaking parameter.

Localisation can be studied through the concept of Lyapunov exponents, which measure the exponential growth rate of the wave function envelope in different channels.
For $\Nc=1$, the Lyapunov exponent of the model can be obtained exactly $\forall\,\varepsilon$ \cite{BouComGeoLeD90,ComTexTou11,GraTexTou14}.
For $\Nc>1$, the determination of the Lyapunov spectrum is extremely complicated in general and is only known explicitly at the symmetry point ($\varepsilon=0$) \cite{BroMudSimAlt98} thanks to an important simplification~\footnote{
  The study of the wave function generally requires to consider two coupled random processes controlling the phase and the envelope of the wave function.
  This analysis simplifies either in the high energy/weak disorder regime when the rapid phase variable decouples from the slow envelope variable, or at $\varepsilon=0$ for the Dirac equation, because the phase variable is locked and the wave function is only controlled by its envelope.
}
(see also Ref.~\cite{LudSchSto13} for a broad perspective on Lyapunov spectra in the various symmetry classes).
Instead of dealing with the FPE~\cite{BroMudSimAlt98,BroMudFur01}, we propose here a more direct derivation of the Lyapunov spectrum from the SDE.

We consider $G=g\mathbf{1}_\Nc$.
Taking our inspiration from the $\Nc=1$ case \cite{RamTex14}, when $\varepsilon=\I k\in\I\mathbb{R}$ we introduce new variables $z_n = k\,\EXP{-2\zeta_n}$, which obey the set of coupled SDEs
$
  \partial_x\zeta_n
  = -k\,\sinh2\zeta_n + \mu\,g
  +\frac{\beta g}{2}\sum_{n(\neq m)}\coth(\zeta_n-\zeta_m) + \tilde{m}_n(x)
$,
where $\tilde{m}_n(x)$ are $\Nc$ \textit{independent} Gaussian white noises of zero mean.
The generator of this diffusion coincides with the one involved in the FPE of Ref.~\cite{BroMudFur01}.
For $\Nc=1$, in the absence of the interaction term, we recover the SDE of \cite{RamTex14,Tex00}. 
For $\varepsilon=\I k=0$, the vanishing of the confinment allows for a simple analysis of the dynamics as the repulsive forces saturate to $\pm1$.
Choosing $\zeta_1<\zeta_2<\cdots<\zeta_\Nc$, one recovers the Lyapunov exponents $\gamma_n = \lim_{x\to\infty}{\zeta_n(x)}/{x} $ of Ref.~\cite{BroMudSimAlt98},~\footnote{
  The Lyapunov spectrum $\gamma_n=g\,(\Nc-2n+1)$ was obtained earlier
  for the matrix $T\exp\big[\int_0^x\D x'\,M(x')\big]$ \cite{New86}, where $T$ is the chronological ordering.
  Thus, \eqref{eq:LyapunovSpectrumZeroEnergy} can be understood from the fact that the zero energy spinor combines the two independent solutions $(1,0)\,T\exp\big[\int_0^x\D x'\,M(x')\big]$ and $(0,1)\,T\exp\big[-\int_0^x\D x'\,M(x')\big]$.
} 
directly from the analysis of the SDEs~:
\begin{equation}
  \label{eq:LyapunovSpectrumZeroEnergy}
  \gamma_n 
  = \Big[ \mu - \frac{\beta}{2}(\Nc-2n+1) \Big]\,g
  \hspace{0.5cm}
  \mbox{for }
  n\in\{1,\cdots,\Nc\}
  \:.
\end{equation}


\section{Density of states}

The density of states (DoS) of multichannel 1D models in the chiral and BdG classes was studied in several papers when $\mu=0$ \cite{BroMudFur00,TitBroFurMud01}~: it was shown that the low energy DoS exhibits a strong dependence in the parity of the channel number $\Nc$.
Whereas the case of even $\Nc$ leads to a low energy DoS 
which depends on the symmetry index $\beta$, for odd $\Nc$ the properties of the strictly 1D Dirac equation with random mass with zero mean were recovered, irrespectively of $\beta$~:
Dyson singularity of the DoS $\rho(\varepsilon)\sim1/\big|\varepsilon\ln^3|\varepsilon|\big|$.
These features were also obtained in two BdG classes with broken spin rotational symmetry  (independently on the parity of $\Nc$) \cite{TitBroFurMud01}.
The existence of common features for distinct symmetry classes was later denoted  as ``\textit{superuniversality}''~\cite{GruReaVis05}, a concept which was used recently for the study of quasi-1D spinless $p$-wave superconducting wires (BdG class D)~\cite{RieBroAda13,RieBro14}.

\begin{figure}[!ht]
\centering
\includegraphics[width=0.45\textwidth]{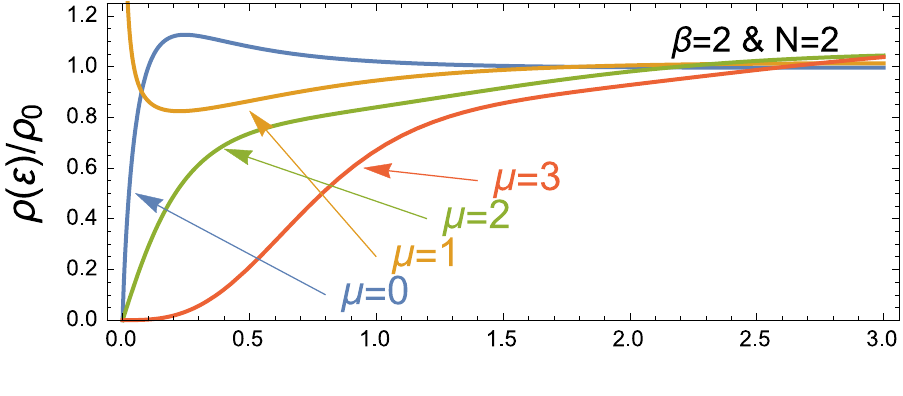}\\[-0.35cm]
\includegraphics[width=0.45\textwidth]{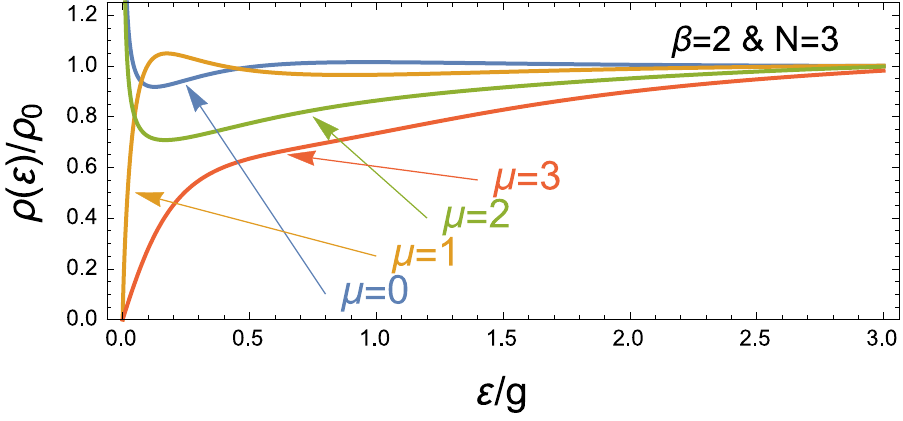}
\caption{\it DoS $\rho(\varepsilon)=\mathcal{N}'(\varepsilon)$ for $\Nc=2$ and $3$ channels for various values of $\mu$ (mean mass over disorder strength). The free DoS is $\rho_0=\Nc/\pi$.}
\label{fig:DoS}
\end{figure}

The analysis of the stochastic process $Z_\varepsilon(x)$ allows for a direct determination of the spectral properties, what was used in the $\Nc=1$ case \cite{BouComGeoLeD90,ComTexTou11,ComLucTexTou13,GraTexTou14}. 
Extending this idea when $\Nc>1$, we introduce the characteristic function $\Omega=\tr{\mean{Z_\varepsilon(x)+M(x)}}$, from which the DoS can be deduced. 
The vanishing or the divergence of $\det Z_\varepsilon(L)$, i.e. of one of its eigenvalues, corresponds to satisfy a second boundary condition for the spinor ($\det\chi(L)=0$ or $\det\varphi(L)=0$) at $x=L$ for one of the eigenmodes.
As a consequence the integrated DoS per unit length is given by $\mathcal{N}(\varepsilon)=-(1/\pi)\im[\Omega]$, as for $\Nc=1$ \cite{ComTexTou11,ComLucTexTou13,GraTexTou14}. 
The mean value $\mean{Z_\varepsilon(x)}$ can be computed from the stationary distribution $f(Z)$, what is more convenient by using that $\Omega$ is an analytic function of $\varepsilon$ and considering purely imaginary energy $\varepsilon=\I k\in\I\mathbb{R}$ because $f(Z)$ 
is an \textit{equilibrium} distribution in this case.
Considering, without loss of generality, $G=\mathrm{diag}\{g_1,\cdots,g_\Nc\}$, we deduce the mean value from the normalisation constant 
$\mean{ \left[Z_\varepsilon(x) + M(x)\right]_{ii} } = g_{i}^2\partial\ln\mathcal{C}_{\Nc,\beta}/\partial{g_i}$.

For the isotropic case, $G=g\,\mathbf{1}_\Nc$, we write equivalently
\begin{equation}
  \label{eq:RelationOmegaNormalisation}
  \Omega 
  = -g\,
  \left(\Nc \, \mu + 
    k  \derivp{\ln\mathcal{C}_{\Nc,\beta}}{k}
  \right)
  \:.
\end{equation}
$\mathcal{C}_{\Nc,\beta}$ can then be written as an integral over the eigenvalues of the Riccati matrix,
$
  \mathcal{C}_{\Nc,\beta}
  =\int_0^\infty\D z_1\cdots\D z_\Nc\,
  \prod_{i<j}|z_i-z_j|^\beta
  \prod_l\phi(z_l)
$
where $\phi(z)=z^{-\mu-1-\beta(\Nc-1)/2}\exp[-(z+k^2/z)/(2g)]$, 
which is computed using standard technique of random matrix theory~\cite{Meh04}. 
The unitary case ($\beta=2$) is the simplest one~: 
we obtain the form of a Hankel determinant
\begin{align}
  \label{eq:CN2}
  \mathcal{C}_{\Nc,2}
  = \Nc! \, 2^\Nc\, k^{-\Nc\mu}\,
    \det\big[K_{\mu+1+\Nc-i-j}(k/g)\big]  
    \:,
\end{align}
where $1\leq i,\,j\leq\Nc$.
$K_\nu(z)$ is the MacDonald function \cite{gragra}.
We can deduce exact expressions for the DoS~: we show in Fig.~\ref{fig:DoS} the DoS for $N=2$ and $N=3$ channels for various values of~$\mu$.
The case $\beta=1$ is discussed in \cite{SM}. 
Assuming that \eqref{eq:StatDistfZ} is also the stationary distribution for $\beta=4$, we get the Pfaffian
\begin{align}
  \label{eq:CN4}
    \mathcal{C}_{\Nc,4} =\Nc! \, 2^\Nc\, k^{-\Nc\mu}\, 
    \mathrm{pf} \big[ (j-i)\,K_{\mu+1+2\Nc-i-j}(k/g) \big] 
  \:,
\end{align}
where $1\leq i,\,j\leq2\Nc$.
For $N=1$, we check that (\ref{eq:CN2},\ref{eq:CN4}) give the known result \cite{BouComGeoLeD90,ComTexTou11}.
Similar results were obtained for~$\mu=0$ by a slightly different approach in Ref.~\cite{TitBroFurMud01}.

\begin{figure}[!ht]
\centering
\includegraphics[height=3.25cm]{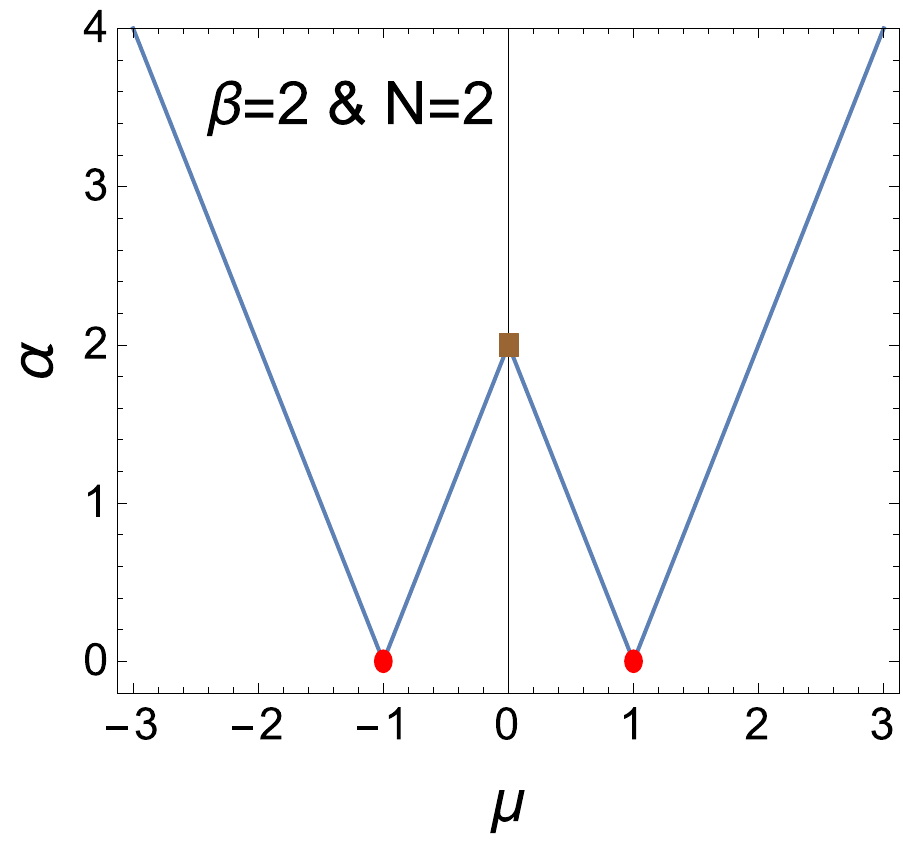}
\hfill
\includegraphics[height=3.25cm]{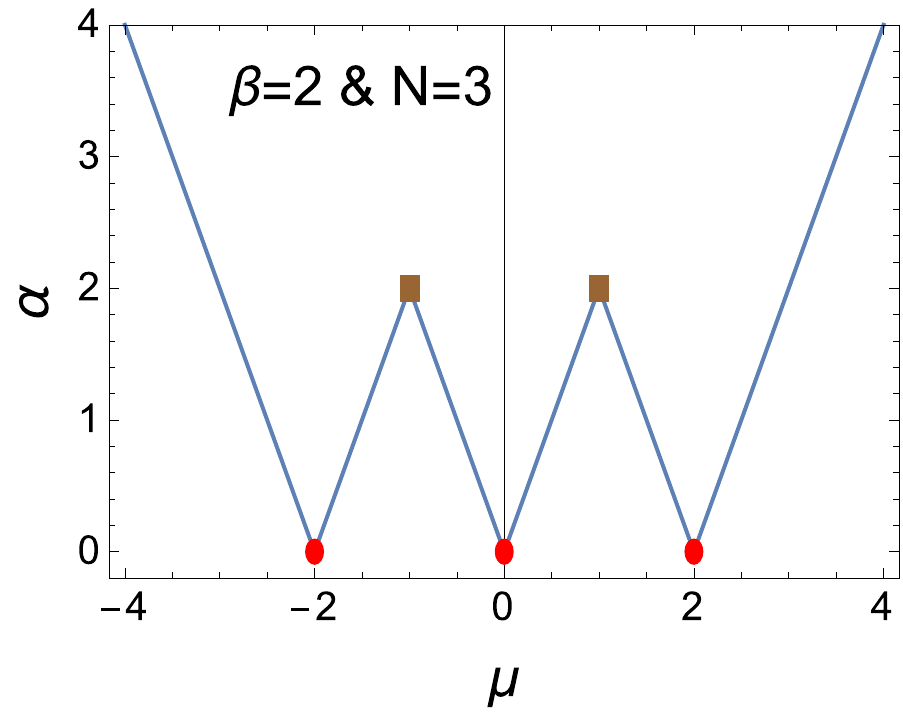}
\caption{\it Exponent controlling the low energy integrated DoS $\mathcal{N}(\varepsilon)\sim\varepsilon^\alpha$ as a function of $\mu=m_0/g$. The dots correspond to the critical points where the IDoS is $\mathcal{N}(\varepsilon)\sim1/\ln^2\varepsilon$. 
The brown squares correspond to the behaviour $\mathcal{N}(\varepsilon)\sim\varepsilon^\beta|\ln\varepsilon|$.}
\label{fig:ExponentAlpha}
\end{figure}

Using \eqref{eq:CN2} or the equivalent expressions for $\beta=1$ and $\beta=4$, we can analyse in detail the low energy behaviour of the DoS, which involves the two leading order terms of a $k\to0$ expansion of $\mathcal{C}_{\Nc,\beta}$.
The DoS presents a power law behaviour $\mathcal{N}(\varepsilon)\sim\varepsilon^\alpha$. 
For $\beta=2$, Eq.~\eqref{eq:CN2} shows that the next leading order term of the $k\to0$ expansion is controlled by an exponent which is a non monotonous function of $\mu$, what originates from the expansion of $K_\nu(z)$. The exponent is $\alpha = 2 \mu - \beta(\Nc-1)$ for $\mu>\beta(\Nc-1)/2$ and vanishes $N$ times below this threshold, presenting a saw behaviour (Fig.~\ref{fig:ExponentAlpha}) (cf.~\cite{SM}). 
Each vanishing of the exponent $\alpha=0$ corresponds to a critical point where the integrated DoS presents the Dyson singularity $\mathcal{N}(\varepsilon)\sim1/\ln^{2}\varepsilon$ (the independence in the Dyson index $\beta$ is a sign of superuniversality \cite{GruReaVis05}).
When the exponent reaches a local maximum (Fig.~\ref{fig:ExponentAlpha}), the behaviour $\mathcal{N}(\varepsilon)\sim\varepsilon^\beta|\ln\varepsilon|$ is found (such a behaviour was found in \cite{BroMudFur00,BroMudFur01} for $\mu=0$ and even~$N$).

\section{Witten index} 

We now demonstrate that the vanishing of the exponent $\alpha$ is a signature of a sequence of $\Nc$ topological phase transitions, accompanied by the change of a quantum number of topological origin.
From the bulk/edge correspondence, we can relate the topological number characterising an insulating phase to the number of zero modes at the boundary of the system.
We consider the Witten index \cite{Wit82}
$\Delta(\tilde\beta) = \mathrm{tr}\{\sigma_3\,\exp[-\tilde\beta\HDirac^2]\}$, 
which can be expressed in terms of the scattering Friedel phase 
$\delta_\pm(\varepsilon)=-\I\ln\det\mathcal{S}_\pm(\varepsilon)$, 
characterising the scattering problem on the half line $x\in\mathbb{R}_+$ for $\mean{M(x)}=\pm m_0\mathbf{1}_N$~\footnote{
    Another connection between topological quantum numbers and scattering was discussed in Ref.~\cite{FulHasAkhBee11}.
}~:
\begin{equation}
  \Delta(\tilde\beta)  
  =\int_0^\infty\D\varepsilon\, \frac{ \delta_+(\varepsilon)-\delta_-(\varepsilon) }{2\pi}\,
  2\tilde\beta\varepsilon\,\EXP{-\tilde\beta\varepsilon^2}
  \:.
\end{equation}
For a continuous spectrum, $\Delta(\tilde\beta)$ presents a non trivial $\tilde\beta$-dependence, however $\Delta(\infty) = \big[\delta_+(0)-\delta_-(0) \big]/(2\pi)$ provides the information on the number of zero modes.

As an elementary illustration, we consider first the free case ($g=0$), when the channels are uncoupled and the spectrum gapped.
Setting $M(x)=-\infty$ (i.e. $\varphi(0)=0$) and $M(x)=m_0\mathbf{1}_\Nc$ on $[0,L]$, we get the reflection phase (for each channel) 
$
\delta(\varepsilon)/2=
\mathrm{arg}
  \big[
    (\kappa\,\cosh\kappa L-m_0\,\sinh\kappa L + \I\,\varepsilon\,\sinh\kappa L)
    \EXP{\I\pi/4}
  \big]
$
where $\kappa=\sqrt{m_0^2-\varepsilon^2}$.
Letting first $L\to\infty$ we deduce that $\lim_{\varepsilon\to0}\delta(\varepsilon)=\pi\,\theta_\mathrm{H}(m_0)+\pi/2$, where $\theta_\mathrm{H}(x)$ is the Heaviside function, leading to $\Delta(\infty)=(\Nc/2)\sign(m_0)$~\footnote{
  For finite $L\gg1/m_0$, the midgap state is splitted into two low energy states, $\varepsilon_\pm\simeq\pm2m_0\exp[-m_0L]$ (for $\varphi(0)=\chi(L)=0$), which is also reflected in the phase shift.
  Distribution of the ground state energy has been obtained in \cite{TexHag10,Tex00}.
}.
Each channel is characterised by a Witten index equal to $1/2$, which is interpreted as a fermion fractionization phenomenon characterising a topologically non trivial state \cite{NieSem86}.
A crucial question is to understand how these states are affected by the disorder.
Quite remarkably the zero modes, which are midgap states in the absence of disordered, exist in strongly disordered wires \cite{RieBroAda13,RieBro14} with \textit{gapless} spectrum, as we show below.


\section{Topological phase transitions}

The spectral density (bulk information) was related to the mean value $\mean{Z_\varepsilon}$~; the distribution $f(Z)$ however contains a lot more of information. 
We now determine the topological index $\Delta(\infty)$ by a direct analysis of \eqref{eq:StatDistfZ}. 
The relation between $\Delta(\infty)$ and $f(Z)$ is made more clear by using the supersymmetry of the Dirac equation~:
denoting $Z_\varepsilon^\pm(x)$ the solution of \eqref{eq:RiccatiEq} for $\mean{M(x)}=\pm\mu G$, we get $Z_\varepsilon^-(x)\eqlaw-\varepsilon^2Z_\varepsilon^+(x)^{-1}$ \cite{ComTexTou11} (equality in law means the same statistical properties).
For $\varepsilon\in\mathbb{R}$ we deduce
$
\mean{\delta_+(\varepsilon)-\delta_-(\varepsilon)}/(2\pi) = 
(1/2)\sum_{n=1}^\Nc\mean{\sign(z_n)}
$. 
For  $\varepsilon\in\I\mathbb{R}$, we get~:
\begin{equation}
  \label{eq:DeltaImaginary}
  \frac{1}{2\pi}
  \mean{\delta_+(\I k)-\delta_-(\I k)} = 
  \frac12
  \sum_{n=1}^\Nc\mean{\sign(z_n-k)}
  \:.
\end{equation}
Below, we will obtain $\Delta(\infty)$ from (\ref{eq:StatDistfZ},\ref{eq:DeltaImaginary}).

\begin{figure}[!ht]
\centering
\includegraphics[height=3.25cm]{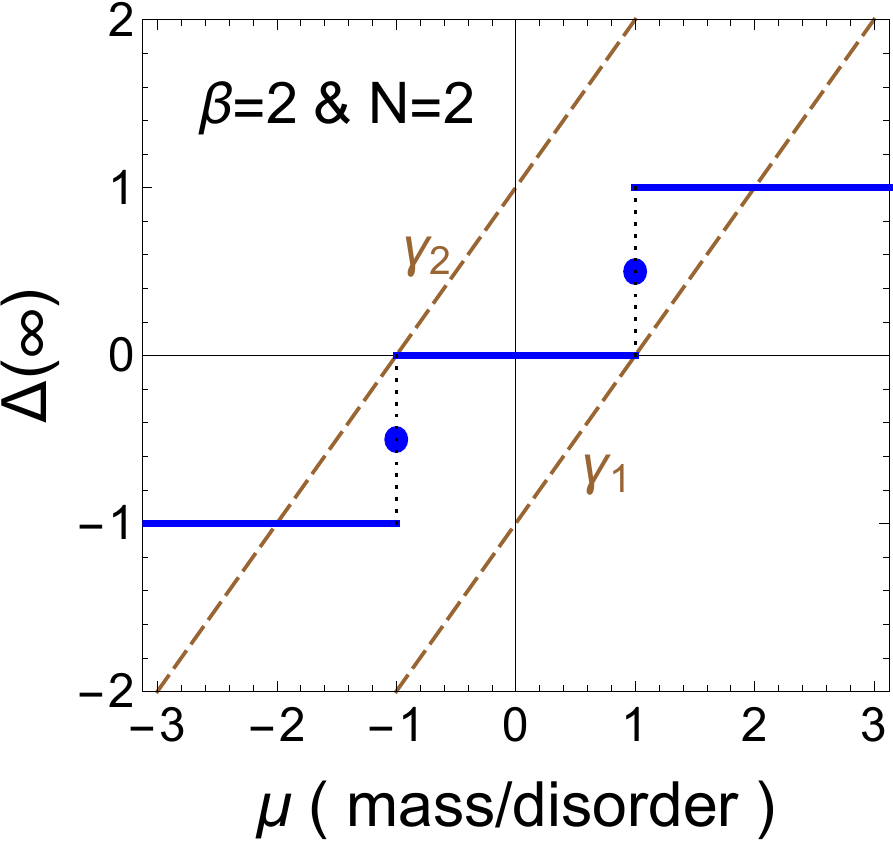}
\hfill
\includegraphics[height=3.25cm]{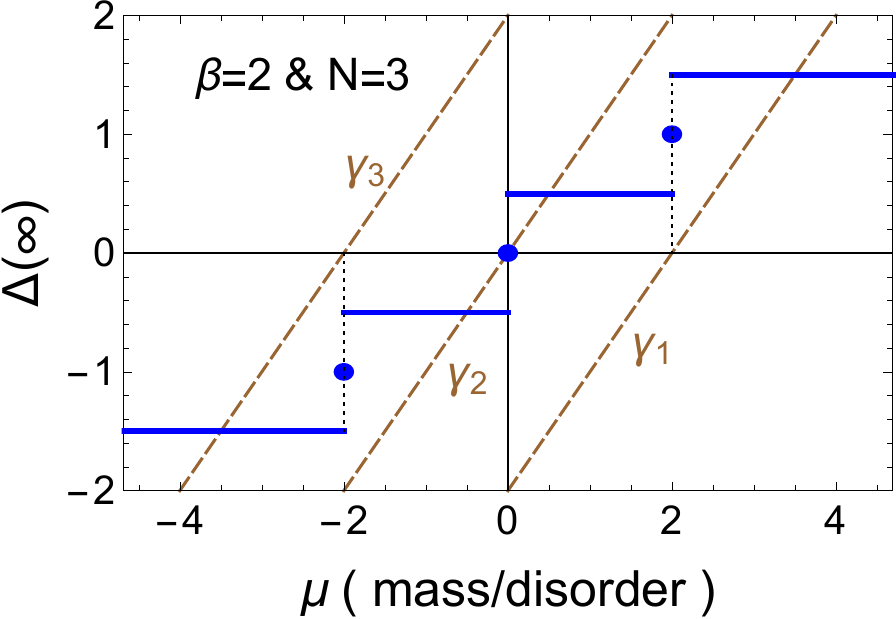}
\caption{\it Witten index $\Delta(\infty)$ (number of zero modes is $2|\Delta(\infty)|$).
  Dashed lines are the Lyapunov exponents. 
  }
  \label{fig:WittenIndex}
\end{figure}

For $G=g\mathbf{1}_\Nc$, the distribution of the Riccati eigenvalues for $\mean{M(x)}=\pm\mu g\mathbf{1}_\Nc$ is 
$P_\pm(z_1,\cdots,z_\Nc)\propto\prod_{i<j}|z_i-z_j|^\beta\prod_l\phi(z_l)$ where $\phi(z)=z^{\mp\mu-1-\beta(\Nc-1)/2}\exp[-(z+k^2/z)/2]$ (setting $g=1$ for simplicity).
In the Coulomb gas approach, $P_\pm$ is interpreted as the Gibbs measure for a 1D gas of $\Nc$ ``charges'' trapped in a confining potential $-\ln\phi(z)$ and with logarithmic interactions.
We show below that the limit $k\to0$ involves the distribution of eigenvalues of the Laguerre ensemble
$\mathcal{L}_{N,\theta}(\lambda_1,\cdots,\lambda_N)
  = A_{N,\theta}^{-1}
  \prod_{i<j} |\lambda_i-\lambda_j|^\beta \,\prod_l \lambda_l^{\theta-1}\,\EXP{-\lambda_l/2}
$,
normalisable for $\theta>0$,
and  the one of the``inverse-Laguerre'' ensemble
$\mathcal{I}_{N,\theta}(\tau_1,\cdots,\tau_N)=A_{N,\theta}^{-1}
  \prod_{i<j} |\tau_i-\tau_j|^\beta \,\prod_l \tau_l^{-\theta-1-\beta(N-1)}\,\EXP{-1/(2\tau_l)}$, i.e. the distribution of $\tau_i=1/\lambda_i$'s.
We first remark that, for $\theta=\mu-\beta(\Nc-1)/2>0$, the limit $k\to0$ of the joint distribution $P_-(z_1,\cdots,z_\Nc)$ corresponds to the Laguerre ensemble. 
Eq.~\eqref{eq:DeltaImaginary} leads to $\Delta(\infty)=\Nc/2$.
For $\tilde{\mu}=\mu-\beta(\Nc-2n-1)/2\in]0,\beta[$ with $n\in\{1,\cdots,\Nc\}$, the limit $k\to0$ is more subtle~:
we obtain that the $\Nc$ charges split into two independent sets, as the distribution behaves as 
\begin{align}
  P_-(z_1, \cdots , z_\Nc)
  \underset{k\to0}{\propto}
  &\frac{1}{k^{2n}}
  \mathcal{I}_{n,\beta-\tilde\mu}\left(\frac{z_1}{k^2}, \cdots , \frac{z_n}{k^2}\right)
  \nonumber\\
  &\times\mathcal{L}_{\Nc-n,\tilde\mu}(z_{n+1}, \cdots , z_\Nc)
  \:.
\end{align}
This distribution describes the ``condensation'' of $n$ charges towards the origin, while the $\Nc-n$ remaining charges are described by a distribution which ``freezes'' as $k\to0$.
Interestingly, the parameter $k$ drives a phase transition in the Coulomb gas (the splitting of the charges into two groups), which occurs for a finite $\Nc$, whereas several phase transitions were observed up to now in the thermodynamic limit $\Nc\to\infty$ in various contexts \cite{VivMajBoh10,TexMaj13,MajSch14}.
As the $n$ smallest charges scale as $k^2\to0$ while the $\Nc-n$ remaining charges scale as $k^0$, we get from Eq.~\eqref{eq:DeltaImaginary} (Fig.~\ref{fig:WittenIndex})~:
\begin{equation}
  \Delta(\infty)= \frac{\Nc-2n}{2}
  \mbox{ for }
  n\in\{0,\,1,\cdots,\,\Nc\}
  \:.
\end{equation}

Exactly at a critical point $\mu=\beta(\Nc-2n+1)/2$ with $n\in\{1,\cdots,\Nc\}$, the charge $n$ must be considered separately~: 
\begin{align}
  &P_-(z_1, \cdots , z_\Nc)
  \underset{k\to0}{\propto}
  \frac{1}{k^{2(n-1)}}
  \mathcal{I}_{n-1,\beta}\left(\frac{z_1}{k^2}, \cdots , \frac{z_{n-1}}{k^2}\right)
  \nonumber\\
  &
  \times
  \frac{1}{z_n}\EXP{-(k/2)[z_n/k+k/z_n]}
  \,
  \mathcal{L}_{\Nc-n,\beta}(z_{n+1}, \cdots , z_\Nc)
  \:.
\end{align}
The position of the charge $n$ now scales as $k\to0$, however its distribution is symmetric with respect to $z_n=k$, which shows that $\mean{\sign(z_n-k)}=0$, thus it does not contribute to \eqref{eq:DeltaImaginary}. As a result  $\Delta(\infty)=(\Nc-2n+1)/2$ (Fig.~\ref{fig:WittenIndex}).
A fine tuning of the disorder (or the mass) can thus change the parity of the number of zero modes.

This analysis can be confronted to the Lyapunov spetrum \eqref{eq:LyapunovSpectrumZeroEnergy} plotted in Fig.~\ref{fig:WittenIndex}~: this suggests that the creation or the annihilation of a pair of zero modes at a topological transition is made possible by the delocalisation in one of the $\Nc$ channels.


\section{Conclusion} 

\begin{figure}[!ht]
\centering
\includegraphics[height=2cm]{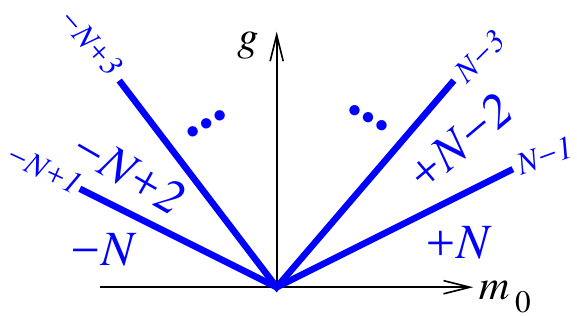}
\caption{\it Phase diagram of the model with $\Nc$ channels~: the half plane (disorder against averaged mass) is splitted in $\Nc+1$ sectors characterised by the value of the topological number $2\Delta(\infty)$. The $\Nc$ lines correspond to $\mu=m_0/g=(\beta/2)(\Nc-2n+1)$, with $n\in\{1,\cdots,\Nc\}$. On each transition line, the index takes the intermediate value.}
  \label{fig:PhaseDiagram}
\end{figure}

In this article we have shown a connection between multichannel 1D disordered models with chiral symmetry and the random matrix model defined by Eq.~\eqref{eq:StatDistfZ}, what has allowed us to derive several exact results straightforwardly (DoS and topological index counting the number of chiral zero modes).
This analysis can be summarized on the phase diagram plotted in Fig.~\ref{fig:PhaseDiagram} (corresponding to the one of Ref.~\cite{MorFurMud15}).
BdG classes D and DIII which support (Majorana) zero modes in 1D \cite{SchRyuFurLud08} have attracted a lot of interest (see \cite{MorFurMud15}), hence it would be interesting to extend our approach to these classes. 

\begin{figure}[!ht]
\centering
\includegraphics[height=3.25cm]{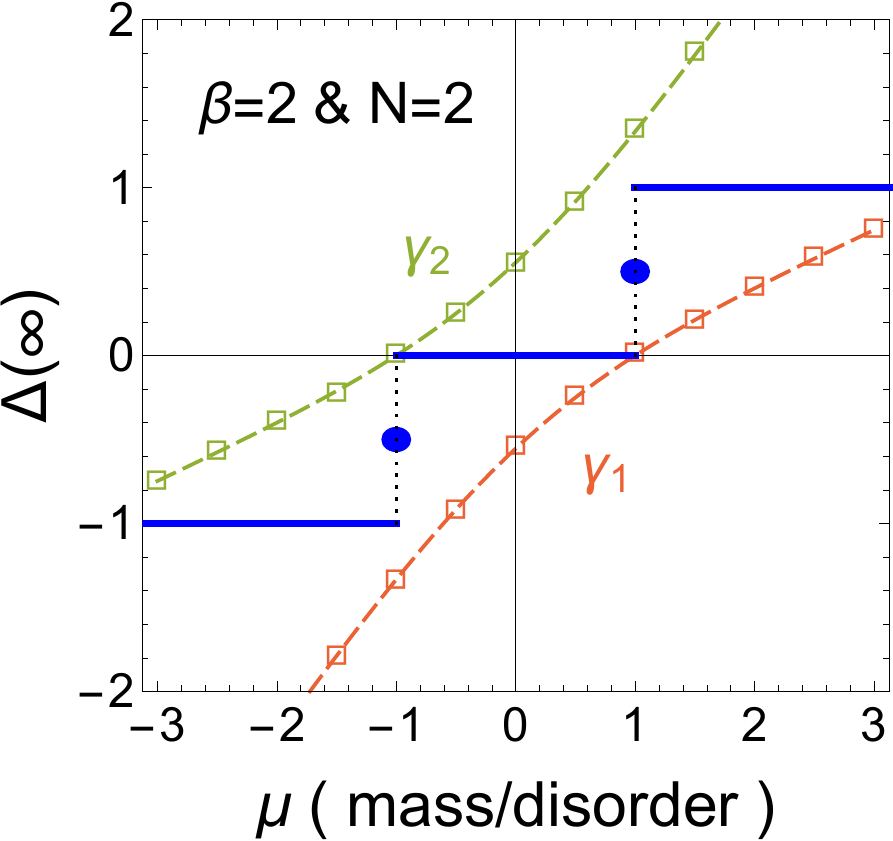}
\hfill
\includegraphics[height=3.25cm]{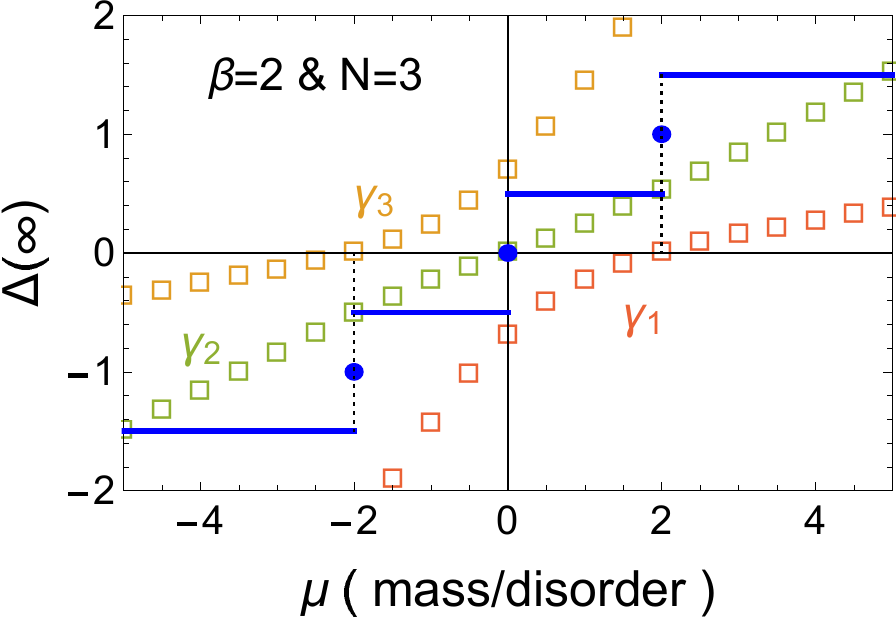}
\caption{\it Witten index $\Delta(\infty)$ and Lyapunov spectra with anisotropic disorder~: for two channels ($g_1/g_2=3$) and three channels ($g_1/g_2=g_2/g_3=3$).
The squares are the values obtained numerically. 
  }
  \label{fig:WittenIndexAnisotrop}
\end{figure}

Although most of the discussion has concerned the isotropic case, the main conclusions can be extended to the non-isotropic case, $G=\mathrm{diag}(g_1,\cdots,g_\Nc)$.
The determinantal representations (\ref{eq:CN2},\ref{eq:CN4}) can be generalised. 
Using these generalisations and the distribution \eqref{eq:StatDistfZ}, we were able to show that both the exponent $\alpha$ and the Witten index present the same dependence with $\mu$ as in the isotropic case (Figs.~\ref{fig:ExponentAlpha} and~\ref{fig:WittenIndex}). 
The Lyapunov analysis is much more difficult to extend.
Using a formalism based on the QR decomposition, we have obtained analytical expressions for the Lyapunov exponents (at $\varepsilon=0$) in the non-isotropic case for $\Nc=2$ channels~: 
\begin{align}
        &\gamma_1 = 2 \frac{g_1 g_2}{g_1+g_2} +
        \mu^2 (g_1 g_2)^\mu \frac{g_1+g_2}{g_1^{2\mu} - g_2^{2\mu}}
        \\\nonumber
              & \times \left[
                        \mathrm{B}\left(
                                \frac{g_1}{g_1+g_2}; 1+\mu, -\mu
                        \right)
                        - \mathrm{B}\left(
                                \frac{g_2}{g_1+g_2}; 1+\mu, -\mu
                        \right)
                \right]
                \\\nonumber
               & +
                \frac{(g_1-g_2)^3}{g_1^{2\mu} - g_2^{2\mu}}
                        \frac{\mu \,g_1^{2\mu}}{6 g_1 g_2}
                        \mathrm{F}_1\! \left(
                                2, 1-\mu, 1+\mu, 4; 1-\frac{g_2}{g_1},1- \frac{g_1}{g_2}
                        \right)
\end{align}
Where $\mathrm{B}(z;a,b)$ is the incomplete Beta-function, and
$\mathrm{F}_1$ is the Appell Hypergeometric function with two arguments:
\begin{equation}
        \mathrm{F}_1 \left(a, b, b', c;x,y \right) =
                \sum_{m \geq 0} \sum_{n \geq 0}
                \frac{(a)_{m+n} (b)_m (b')_n}{(c)_{m+n} m! n!} x^m y^n,
\end{equation}
with $(a)_n = \Gamma(a+n)/\Gamma(a)$. 
The second Lyapunov exponent follows from the  sum rule $\sum_n\gamma_n=\mu\,\tr{G}$ (valid at $\varepsilon=0$), i.e. $\gamma_2 = \mu\,(g_1+g_2) - \gamma_1$.
For $\Nc>2$, we have performed a numerical analysis. 
Our results also exhibit the vanishing of one Lyapunov at each phase transition, as expected (Fig.~\ref{fig:WittenIndexAnisotrop}). 
Details will be published elsewhere.

Although we are providing here the first exact results for a non-isotropic multichannel disorder, to the best of our knowledge, we stress that we have not considered here the most general situation, which would involve the distribution of disorder of the form $P[M(x)]\propto\exp\big[-\int\D x\,\tr{A^{-1}M^\dagger(x)B^{-1}M(x)}\big]$. This remains a challenging problem.


\section{Acknowledgements}
We acknowledge numerous stimulating discussions with Alain Comtet and Yves Tourigny. 
CT thanks Fr\'ed\'eric Pi\'echon and Gr\'egory Schehr for useful remarks.


\begin{thebibliography}{40}

\bibitem{Bee97}
\Name{Beenakker C. W.~J.}
Random-matrix theory of quantum transport, 
\REVIEW{Rev. Mod. Phys.}{69}{1997}{731}.

\bibitem{MelKum04}
\Name{Mello P.~A. \and Kumar N.} 
\Book{Quantum transport in mesoscopic systems -- Complexity and statistical fluctuations} 
(Oxford University Press) 2004.

\bibitem{Zir96}
\Name{Zirnbauer M.~R.}
  Riemannian symmetric superspaces and their origin in random-matrix theory, 
  \REVIEW{J. Math. Phys.}{37}{1996}{4986}.

\bibitem{AltZir97}
\Name{Altland A. \and Zirnbauer M.~R.}
 Nonstandard symmetry classes in mesoscopic normal-superconducting hybrid structures,
\REVIEW{Phys. Rev.~B}{55}{1997}{1142}.

\bibitem{EveMir08}
\Name{Evers F. \and Mirlin A.~D.}
 Anderson transitions,
\REVIEW{Rev. Mod. Phys.}{80}{2008}{1355}.

\bibitem{BroMudSimAlt98}
\Name{Brouwer P.~W., Mudry C., Simons B.~D. \and Altland A.}
 Delocalization in Coupled One-Dimensional Chains,
\REVIEW{Phys. Rev.Lett.}{81}{1998}{862}.

\bibitem{BroMudFur00}
\Name{Brouwer P.~W., Mudry C. \and Furusaki A.}
 Density of States in Coupled Chains with Off-Diagonal Disorder,
\REVIEW{Phys. Rev. Lett.}{84}{2000}{2913}.

\bibitem{BroMudFur01}
\Name{Brouwer P.~W., Mudry C. \and Furusaki A.}
 Transport properties and density of States of quantum wires with
  Off-Diagonal Disorder,
\REVIEW{Physica E}{9}{2001}{333}.

\bibitem{TitBroFurMud01}
\Name{Titov M., Brouwer P.~W., Furusaki A. \and Mudry C.}
 Fokker-Planck equations and density of states in disordered quantum
  wires,
\REVIEW{Phys. Rev. B}{63}{2001}{235318}.

\bibitem{Kit01}
\Name{\relax{Yu}. Kitaev A.}
 Unpaired Majorana fermions in quantum wires,
\REVIEW{Phys. Usp.}{44}{2001}{131}.

\bibitem{OreRefOpp10}
\Name{Oreg Y., Refael G. \and von Oppen F.}
 Helical Liquids and Majorana Bound States in Quantum Wires,
\REVIEW{Phys. Rev. Lett.}{105}{2010}{177002}.

\bibitem{AliOreRefOppFis11}
\Name{Alicea J., Oreg Y., Refael G., von Oppen F. \and Fisher M. P.~A.}
 Non-Abelian statistics and topological quantum information processing
  in 1D wire networks,
  \REVIEW{Nature Physics}{7}{2011}{412}.

\bibitem{MotDamHus01}
\Name{Motrunich O., Damle K. \and Huse D.~A.}
 Griffiths effects and quantum critical points in dirty
  superconductors without spin-rotation invariance: One-dimensional examples,
\REVIEW{Phys. Rev. B}{63}{2001}{224204}.

\bibitem{PotLee10}
\Name{Potter A.~C. \and Lee P.~A.}
 Multichannel Generalization of Kitaev's Majorana End States and a
  Practical Route to Realize Them in Thin Films,
\REVIEW{Phys. Rev. Lett.}{105}{2010}{227003}.

\bibitem{RieBroAda13}
\Name{Rieder M.-T., Brouwer P.~W. \and Adagideli I.}
 Reentrant topological phase transitions in a disordered spinless
  superconducting wire,
 \REVIEW{Phys. Rev. B}{88}{2013}{060509}.

\bibitem{LudSchSto13}
\Name{Ludwig A. W.~W., Schulz-Baldes H. \and Stolz M.}
 Lyapunov Spectra for All Ten Symmetry Classes of
  Quasi-one-dimensional Disordered Systems of Non-interacting Fermions,
 \REVIEW{J. Stat. Phys.}{152}{2013}{275}.

\bibitem{MorFurMud15}
\Name{Morimoto T., Furusaki A. \and Mudry C.}
 Anderson localization and the topology of classifying spaces,
 \REVIEW{Phys. Rev. B}{91}{2015}{235111}.

\bibitem{RieBro14}
\Name{Rieder M.-T. \and Brouwer P.~W.}
 Density of states at disorder-induced phase transitions in a
  multichannel Majorana wire,
 \REVIEW{Phys. Rev. B}{90}{2014}{205404}.

\bibitem{Bee15}
\Name{Beenakker C. W.~J.}
 Random-matrix theory of Majorana fermions and topological
  superconductors,
 \REVIEW{Rev. Mod. Phys.}{87}{2015}{1037}.

\bibitem{BouComGeoLeD90}
\Name{Bouchaud J.-P., Comtet A., Georges A. \and {Le~Doussal} P.}
 Classical diffusion of a particle in a one-dimensional random force
  field,
\REVIEW{Ann. Phys. (N.Y.)}{201}{1990}{285}.

\bibitem{ComTexTou11}
\Name{Comtet A., Texier C. \and Tourigny Y.}
 Supersymmetric quantum mechanics with L\'{e}vy disorder in one
  dimension,
 \REVIEW{J. Stat. Phys.}{145}{2011}{1291}.

\bibitem{ComLucTexTou13}
\Name{Comtet A., Luck J.-M., Texier C. \and Tourigny Y.}
 The Lyapunov exponent of products of random $2\times2$ matrices close
  to the identity,
 \REVIEW{J. Stat. Phys.}{150}{2013}{13}.

\bibitem{GraTexTou14}
\Name{Grabsch A., Texier C. \and Tourigny Y.}
 One-dimensional disordered quantum mechanics and Sinai diffusion with
  random absorbers,
 \REVIEW{J. Stat. Phys.}{155}{2014}{237}.

\bibitem{AntComKne90}
\Name{Antoine M., Comtet A. \and Knecht M.}
 Heat Kernel expansion for fermionic billiards in an external magnetic
  field,
 \REVIEW{J.~Phys.~A: Math. Gen.}{23}{1990}{L35}.

\bibitem{TexHag10}
\Name{Texier C. \and Hagendorf C.}
 Effect of boundaries on the spectrum of a one-dimensional random mass
  Dirac Hamiltonian,
 \REVIEW{J.~Phys.~A: Math. Theor.}{43}{2010}{025002}.

\bibitem{Tex00}
\Name{Texier C.}
  Individual energy level distributions for one-dimensional diagonal and off-diagonal disorder,
 \REVIEW{J.~Phys.~A: Math. Gen.}{33}{2000}{6095}.

\bibitem{RamTex14}
\Name{Ramola K. \and Texier C.}
 Fluctuations of random matrix products and 1D Dirac equation with
  random mass,
 \REVIEW{J. Stat. Phys.}{157}{2014}{497}.

\bibitem{New86}
\Name{Newman C.~M.}
 The distribution of Lyapunov exponents: Exact results for random
  matrices,
 \REVIEW{Commun. Math. Phys.}{103}{1986}{121}.

\bibitem{GruReaVis05}
\Name{Gruzberg I.~A., Read N. \and Vishveshwara S.}
 Localization in disordered superconducting wires with broken
  spin-rotation symmetry,
 \REVIEW{Phys. Rev. B}{71}{2005}{245124}.

\bibitem{Meh04}
\Name{Mehta M.~L.} 
\Book{Random matrices} 
3rd Edition (Elsevier, Academic, New York) 2004.

\bibitem{gragra}
\Name{Gradshteyn I.~S. \and Ryzhik I.~M.} 
\Book{Table of integrals, series and products} 
5th Edition (Academic Press) 1994.

\bibitem{Wit82}
\Name{Witten E.}
 Constraints on supersymmetry breaking,
\REVIEW{Nucl. Phys. B}{202}{1982}{253}.

\bibitem{FulHasAkhBee11}
\Name{Fulga I.~C., Hassler F., Akhmerov A.~R. \and Beenakker C. W.~J.}
 Scattering formula for the topological quantum number of a disordered
  multimode wire,
  \REVIEW{Phys. Rev. B}{83}{2011}{155429}.

\bibitem{NieSem86}
\Name{Niemi A.~J. \and Semenoff G.~W.}
 Fermion number fractionization in quantum field theory,
 \REVIEW{Phys. Rep.}{135}{1986}{99}.

\bibitem{VivMajBoh10}
\Name{Vivo P., Majumdar S.~N. \and Bohigas O.}
 Probability distributions of linear statistics in chaotic cavities
  and associated phase transitions,
 \REVIEW{Phys. Rev. B}{81}{2010}{104202}.

\bibitem{TexMaj13}
\Name{Texier C. \and Majumdar S.~N.}
 Wigner time-delay distribution in chaotic cavities and freezing
  transition,
 \REVIEW{Phys. Rev. Lett.}{110}{2013}{250602} erratum: {\it ibid} {\bf112}, 139902 (2014).

\bibitem{MajSch14}
\Name{Majumdar S.~N. \and Schehr G.}
 Top eigenvalue of a random matrix: large deviations and third order
  phase transition,
 \REVIEW{J. Stat. Mech.}{}{2014}{P01012}.

\bibitem{SchRyuFurLud08}
\Name{Schnyder A.~P., Ryu S., Furusaki A. \and Ludwig A. W.~W.}
 Classification of topological insulators and superconductors in three
  spatial dimensions,
 \REVIEW{Phys.  Rev. B}{78}{2008}{195125}.

\bibitem{SM}
  See supplemental material at [URL inserted by EPL].

\end{thebibliography}

\begin{thebibliography}{10}

\bibitem[EM08]{EveMir08SM}
F.~Evers and A.~D. Mirlin,
 Anderson transitions,
 Rev. Mod. Phys. {\bf 80}(4), 1355--1417 (2008).

\bibitem[M82]{Mui82SM}
R.~J. Muirhead,
 {\em Aspects of multivariate statistical theory},
 Wiley, New York, 1982.

\bibitem[BCGL90]{BouComGeoLeD90SM}
J.-P. Bouchaud, A.~Comtet, A.~Georges, and P.~{Le~Doussal},
 Classical diffusion of a particle in a one-dimensional random force
  field,
 Ann. Phys. (N.Y.) {\bf 201}, 285--341 (1990).

\bibitem[CTT11]{ComTexTou11SM}
A.~Comtet, C.~Texier, and Y.~Tourigny,
 Supersymmetric quantum mechanics with L\'{e}vy disorder in one
  dimension,
 J. Stat. Phys. {\bf 145}(5), 1291--1323 (2011).

\bibitem[PP12]{PedPed12SM}
K.~B. Pedersen and M.~S. Pedersen,
 {\em The matrix cookbook},
 Technical University of Denmark, 2012.

\bibitem[M04]{Meh04SM}
M.~L. Mehta,
 {\em Random matrices},
 Elsevier, Academic, New York, third edition, 2004.

\bibitem[B56]{deB56SM}
N.~G. {de Bruijn},
 On some multiple integrals involving determinants,
 J. Indian Math. Soc. {\bf 19}, 133--151 (1956).

\bibitem[SCFG99]{SteCheFabGog99SM}
M.~Steiner, Y.~Chen, M.~Fabrizio, and A.~O. Gogolin,
 Statistical properties of localization-delocalization transition in
  one dimension,
 Phys. Rev.~B {\bf 59}(23), 14848--14851 (1999).

\bibitem[BFGM00]{BroFurGruMud00SM}
P.~W. Brouwer, A.~Furusaki, I.~A. Gruzberg, and C.~Mudry,
 Localization and Delocalization in Dirty Superconducting Wires,
 Phys. Rev. Lett. {\bf 85}(5), 1064 (2000).

\bibitem[BMF01]{BroMudFur01SM}
P.~W. Brouwer, C.~Mudry, and A.~Furusaki,
 Transport properties and density of States of quantum wires with
  Off-Diagonal Disorder,
 Physica E {\bf 9}, 333--339 (2001).

\bibitem[RM14]{RamTex14SM}
K.~Ramola and C.~Texier,
 Fluctuations of random matrix products and 1D Dirac equation with
  random mass,
 J. Stat. Phys. {\bf 157}(3), 497--514 (2014).



\end{thebibliography}


\newpage
\ 
\newpage

\begin{center}
{\bf\large Supplementary material}

\vspace{0.25cm}

{Aur\'elien Grabsch and Christophe Texier}

\vspace{0.25cm}
\end{center}

\noindent
\textbf{Abstract -- }
{\small We present some technical details~:
(\textit{i}) proof of the stationary matrix distribution (main result of the letter, Eq.~1).
(\textit{ii}) Derivation of the exponent of the low energy DoS (the orthogonal and symplectic cases are also discussed).
Finally we discuss the relation between the Lyapunov analysis and the conductance.
}

\vspace{0.25cm}

\section{A. The Fokker-Planck equation for the matricial process}


Let us write the Fokker-Planck equation associated to the SDE satisfied by the Riccati matrix~:
\begin{equation}
	Z' = - Z^2 - E - \mu G Z - \mu Z G - Z\widetilde{M} - \widetilde{M}Z,
\end{equation}
where $E=\varepsilon^2$ and $\widetilde{M}(x)$ is a Gaussian white noise distributed according to~:
\begin{equation}
	P[\widetilde{M}(x)] \propto \exp \left[ - \frac{1}{2} \int \D x \,
		 \tr{  \widetilde{M}(x)^\dagger G^{-1} \widetilde{M}(x) } \right],
\end{equation}
with $\widetilde{M}$ a Hermitian $\Nc\times\Nc$ matrix with real ($\beta=1$), complex ($\beta=2$) or quaternionic ($\beta=4$) elements. The Hermitian matrix $G$ controls the correlations. 
The three cases correspond respectively to the classes BDI (chiral orthogonal), AIII (chiral unitary) and CII (chiral symplectic) of the classification of disordered systems (cf. \cite{EveMir08SM}).
One can assume without loss of generality that $G$ is diagonal
$G=\mathrm{diag}(g_1,\cdots,g_N)$. 
This implies that the entries of $\widetilde{M}$ are independent Gaussian white noises. The number of independent components of $\widetilde{M}$ (and also of $Z$) being
\begin{equation}
	\beta \frac{N(N-1)}{2} + N,
\end{equation}
one has to distinguish the cases $\beta=1$ and $\beta=2$ in the derivation of the Fokker-Planck equation.

\subsection{A.1. Case $\beta = 1$ (class BDI)}

In this case the matrices $\widetilde{M}$ and $Z$ are real and symmetric, so one has to consider only their upper-triangular part. Denote
\begin{equation}
	\widetilde{M}_{ab} = \sqrt{\sigma_{ab}} \zeta_{ab},
\end{equation}
where $\zeta_{ab}$ is a Gaussian white noise of variance $1$ and
\begin{equation}
	\sigma_{ab} = \left\lbrace
		\begin{array}{c l}
			g_a & \text{if } a=b,\\
			\frac{g_a g_b}{g_a + g_b} & \text{if } a \neq b.
		\end{array}
	\right.
	\label{eq_variance}
\end{equation}
Writing the SDE in components
\begin{equation}
	Z_{mn}' = \left[ - Z^2 - E - \mu G Z - \mu Z G  \right]_{mn}
		+ \sum_{k \leq l} B_{mn,kl}(Z) \zeta_{kl},
\end{equation}
where
\begin{align}
	\label{eq_B}
	B_{mn,kl}(Z) &= -\frac{2-\delta_{kl}}{2} \sqrt{\sigma_{kl}}
	\\ \nonumber
	 &\times\left( Z_{mk} \delta_{nl} + Z_{ml} \delta_{nk}
			+ Z_{ln} \delta_{km} + Z_{kn} \delta_{lm} \right)
     \:.
\end{align}
One can now write the adjoint generator of the diffusion as~:
\begin{align*}
	\mathscr{G}^\dagger &= 
	\sum_{m \leq n} \derivp{}{Z_{mn}} \left[  Z^2 + E + \mu G Z + \mu Z G  \right]_{mn}\\
		&+ \frac{1}{2} \sum_{m \leq n} \sum_{k \leq l} \sum_{a \leq b}
			\derivp{}{Z_{mn}}  B_{mn,kl}(Z) \derivp{}{Z_{ab}} B_{ab,kl}(Z).
\end{align*}
To take into account the symmetry of $Z$, introduce the differential operator \cite{Mui82SM}~:
\begin{equation}
	\left( \derivp{}{Z} \right)_{mn} = \frac{1+\delta_{mn}}{2} \derivp{}{Z_{mn}},
\end{equation}
and also a modified matrix of variances~:
\begin{equation}
	\tilde{\sigma}_{ab} = \frac{2-\delta_{ab}}{2} \sigma_{ab}.
	\label{eq_sigmatilde}
\end{equation}
The generator can then be expressed in the compact form~:
\begin{align*}
	\mathscr{G}^\dagger =
	2 \,& \tr{ 
		\derivp{}{Z} Z \left[
			\tilde{\sigma} \circ \left[ \derivp{}{Z} Z + \left( \derivp{}{Z} Z \right)^T \right]
		\right]
	} \\
	&+ \tr{ \derivp{}{Z} \left( Z^2 + E + \mu G Z + \mu Z G \right) },
\end{align*}
where $[A \circ B]_{mn} = A_{mn} B_{mn}$ is the Hadamard product.

\subsection{A.2. Case $\beta = 2$ (class AIII)}

In this case, one can treat the non diagonal entries $Z_{mn}$ and $Z_{nm}$ independently. But now, the non diagonal entries of $\widetilde{M}$ are complex. Denote~:
\begin{equation}
	\widetilde{M}_{mn} = \sqrt{\sigma_{mn}} (\zeta_{mn} + i \xi_{mn}),
\end{equation}
where $\sigma$ is given by \eqref{eq_variance}, $\zeta_{mn}$ and $\xi_{mn}$ are Gaussian white noises of variance $1$, with $\zeta_{mn} = \zeta_{nm}$ and $\xi_{mn} = -\xi_{nm}$. The SDE gives, in components~:
\begin{align}
	Z_{mn}' &= \left[ - Z^2 - E - \mu G Z - \mu Z G  \right]_{mn}
	\nonumber\\
		&+ \sum_{k \leq l} B_{mn,kl}(Z) \, \zeta_{kl}
		+ \sum_{k < l} C_{mn,kl}(Z) \, \xi_{kl},
\end{align}
with $B(Z)$ given by \eqref{eq_B}, and
\begin{align}
	C_{mn,kl}(Z) = &- \sqrt{\sigma_{kl}}
	\\\nonumber
	&\times
	 \left( Z_{mk} \delta_{nl} - Z_{ml} \delta_{nk} 
			+ Z_{ln} \delta_{km} - Z_{kn} \delta_{lm} \right).
\end{align}
The generator is now~:
\begin{align*}
	\mathscr{G}^\dagger = 
	\sum_{m, n} \derivp{}{Z_{mn}} \left[  Z^2 + E + \mu G Z + \mu Z G  \right]_{mn}\\
		+ \frac{1}{2} \sum_{m, n} \sum_{k \leq l} \sum_{a, b}
			\derivp{}{Z_{mn}}  B_{mn,kl}(Z) \derivp{}{Z_{ab}} B_{ab,kl}(Z)\\
	+  \frac{1}{2} \sum_{m, n} \sum_{k < l} \sum_{a, b}
			\derivp{}{Z_{mn}}  C_{mn,kl}(Z) \derivp{}{Z_{ab}} C_{ab,kl}(Z).
\end{align*}
Denoting simply
\begin{equation}
	\left( \derivp{}{Z} \right)_{mn} =  \derivp{}{Z_{mn}},
\end{equation}
and again $\tilde{\sigma}$ defined by \eqref{eq_sigmatilde}, the generator takes the form~:
\begin{align*}
&
	\mathscr{G}^\dagger =
	\mathrm{tr}\bigg\{
	\\
&
		\left[
			\derivp{}{Z} Z^T + \left( \left( \derivp{}{Z} \right)^T  \!\!\! Z \right)^T
		\right]^T
		\!\!\!
		\tilde{\sigma} 
		\circ
		\left[
			\left( \derivp{}{Z} \right)^T  \!\!\!  Z + \left( \derivp{}{Z} Z^T \right)^T
		\right]	 
		\\
&		
	+ 
		\left( \derivp{}{Z} \right)^T
		\left[
			Z^2 + E + \mu G Z + \mu Z G
		\right]
	\bigg\}
\end{align*}
 
\subsection{A.3. Stationary solution}

The well-known form for the stationary solution in the case $N=1$ \cite{BouComGeoLeD90SM,ComTexTou11SM} has led us to propose the distribution
\begin{align}
  \label{eq:SS-SM}
	f(Z) = &\:\mathcal{C}_{N,\beta}^{-1} \det(Z)^{-\mu - 1 - \beta(N-1)/2} 
	\nonumber\\
	 & \times \exp \left[ -\frac{1}{2} \tr{ G^{-1} (Z - E Z^{-1}) } \right]
	 \:.
\end{align}
Using the expression of the generators given above and standard formulae for derivatives of traces and determinants \cite{PedPed12SM}, we have verified after lengthy calculations that \eqref{eq:SS-SM} is a solution of the stationary Fokker-Planck equation~:
\begin{equation}
	\mathscr{G}^\dagger f = 0
	\:.
\end{equation}


\section{B. DoS in the isotropic case}

When the correlation matrix $G$ is proportional to unity, $G=g\mathbf{1}_N$, the normalisation constant reduces to a multiple integral over the $N$ eigenvalues of the Riccati matrix~:
\begin{equation}
  \label{eq:IntegralCN}
  \mathcal{C}_{\Nc,\beta} = 
  \int_0^\infty\D z_1\cdots\D z_\Nc\,
  \prod_{i<j}|z_i-z_j|^\beta
  \prod_l\phi(z_l)
  \:,
\end{equation}
where $\phi(z)=z^{-\mu-1-\beta(\Nc-1)/2}\exp[-(z+k^2/z)/(2g)]$. Such integrals may be written under the form of determinants thanks to standard technique~\cite{Meh04SM}.

The low energy behaviour of the DoS can be obtained from the two leading orders of a $k \to 0$ expansion of $\mathcal{C}_{N,\beta}$~:
\begin{equation}
	\mathcal{C}_{N,\beta} \sim k^\eta ( 1 + A k^\alpha + o(k^\alpha)).
\end{equation}
Using Eq.~(\textbf{6}) of the letter, one gets the expansion of the characteristic function~:
\begin{equation}
	\Omega = - N \mu g - \eta g - \alpha A g k^\alpha + o(k^\alpha).
\end{equation}
Then, analytic continuation to $k = - \I \varepsilon$ yields $\mathcal{N}(\varepsilon) \sim \varepsilon^\alpha$.
We now discuss how the exponent $\alpha$ can be determined.

\subsection{B.1. Case $\beta=2$ (class AIII)}

The unitary case is the most simple one~: the integral \eqref{eq:IntegralCN} may be rewritten under the form of a determinant $\mathcal{C}_{\Nc,2}=N!\,\det(A)$, where the $N\times N$ matrix $A$ has elements 
\begin{equation}
  A_{i,j} = \int_0^\infty\D z\, \phi(z)\, z^{i+j-2}
  \:.
\end{equation}
We recognize the MacDonald function, hence
\begin{align}
  \label{eq:CN2-SM}
  \hspace{-0.05cm}
  \mathcal{C}_{\Nc,2}
  = \Nc! \, 2^\Nc\, k^{-\Nc\mu}
    \det\big[K_{\mu+1+\Nc-i-j}(k/g)\big]_{1\leq i,\,j\leq\Nc}
    \:.
\end{align}

\vspace{0.125cm} 

\noindent $\bullet$ \textit{Low energy analsyis.---}
The analysis of the $k \to 0$ behaviour of the normalisation constant relies on the properties of the MacDonald functions~:
\begin{equation}
	K_\nu(k) \sim \left\lbrace
		\begin{array}{ll}
			k^{- \abs{\nu}} ( 1 + A_\nu k^2 + o(k^2)) & \text{ for } \abs{\nu} \geq 1,\\
			k^{- \abs{\nu}} (1 + B_\nu k^{2 \abs{\nu}} + o(k^{2 \abs{\nu}})) & \text{ for } \abs{\nu}<1.
		\end{array}
		\right.		
\end{equation}
Introduce 
\begin{equation}
 \nu = \mu - N + n \in ]0,1[
  \:,
\end{equation} 
with $n \in \{  1, \ldots ,N \}$. 
In the expansion of the determinant, many terms give a contribution to the two leading orders. To get the power of $k$, one only needs to exhibit one of these terms. For example, the product of all diagonal terms gives a contribution to the leading order , giving a behaviour $k^\eta$ (the exact expression of $\eta$ is not important). To get the next leading term, one has to distinguish two cases depending on the parity of $n$.

\vspace{0.125cm} 

\noindent $\bullet$ \textit{Odd $n$.---}
On the diagonal, the MacDonald function with the smallest index is $K_{\nu}$, which gives a contribution $k^{-\nu} (1 + B_\nu k^{2\nu})$. This gives the expansion $\mathcal{C}_{N,\beta} \sim k^\eta ( 1 + A k^{2 \nu})$, from which one gets
\begin{equation}
  \label{eq:Saw1}
	\alpha = 2 \nu = 2(\mu - N + n)
	\:.
\end{equation}

\vspace{0.125cm} 

\noindent $\bullet$ \textit{Even $n$.---}
Now, the MacDonald function with the smallest index is $K_{\nu-1}$. Proceeding as before, it gives $\mathcal{C}_{N,\beta} \sim k^\eta (1 + B k^{2(1-\nu)})$, so~:
\begin{equation}
  \label{eq:Saw2}
	\alpha = 2 (1-\nu) = 2 (N - n + 1 - \mu)
	\:.
\end{equation}

The two behaviours (\ref{eq:Saw1},\ref{eq:Saw2}) correspond to the saw behaviour shown in Fig.~\textbf{2} of the letter.

\subsection{B.2. Case $\beta=1$ (class BDI)}

We wish to calculate the matrix integral \eqref{eq:IntegralCN} for $\beta=1$.
The case of even and odd $\Nc$ must be treated separately.

\vspace{0.125cm} 

\noindent $\bullet$ \textit{Even $\Nc$.---}
We use a standard result due to de Bruijn~\cite{deB56SM} which allows to express the normalisation in terms of a Pfaffian
\begin{equation}
  \label{eq:PfaffianRepres}
  \mathcal{C}_{\Nc,1} =  \Nc!  \, \pf( A ) 
\end{equation}
where $A$ is the $N\times N$ antisymmetric matrix with elements
\begin{equation}
  \label{eq:AntisymMatrixA}
  A_{i,j}
  =\int_0^\infty\D z_1\D z_2\,
  \sign(z_1-z_2)\,z_1^{i-1}z_2^{j-1}\phi(z_1)\phi(z_2)
  \:.
\end{equation}

\vspace{0.125cm} 

\noindent $\bullet$ \textit{Odd $\Nc$.---}
Eq.~\eqref{eq:PfaffianRepres} holds provided one adds a line and a column to the matrix $A$, given by~:
\begin{align}
  \label{eq:AntisymMatrixAOddN}
	A_{i,N+1} &= - A_{N+1,i} = \int_0^\infty \D z\, z^{i-1} \phi(z)
	\\
		&= 2 k^{\mu + i - \frac{N+1}{2}} K_{\frac{N+1}{2} - \mu - i}(k)
		\:.
\end{align}

\vspace{0.125cm} 

\noindent $\bullet$ \textit{Low energy analysis.---}
The leading order as $k \to 0$ can be obtained directly from \eqref{eq:AntisymMatrixA} by rescaling either $z_1$ or $z_2$ by a factor $k^2$ in order to obtain convergent integrals in the limit $k \to 0$. 
To get the next order, we rescale the variables as $z_1= k x$ and $z_2= k y$. 
This gives~:
\begin{equation}
	A_{i,j} = k^{i+j+2\mu -N-1} B_{i+\mu-\frac{N+1}{2}, j+\mu-\frac{N+1}{2}}(k),
\end{equation}
where we have introduced~:
\begin{align}
	B_{\alpha,\beta}(k) = \int_0^\infty \D x \int_0^\infty \D y \, 
	  &\sign(x-y)
	  \\\nonumber
	 &\times x^{\alpha-1} y^{\beta-1} e^{-\frac{k}{2} \left( x + y + {1}/{x} + {1}/{y} \right)}
	 \:.
\end{align}
From this expression, it can be shown that
\begin{align}
	(\alpha+\beta) B_{\alpha,\beta}(k) + k \partial_k B_{\alpha,\beta}(k) =
		&- k B_{\alpha-1,\beta}(k) 
		\nonumber\\	
	& - k B_{\alpha,\beta-1}(k) 
		\:,
    \\		
	(\alpha-\beta) B_{\alpha,\beta}(k) + k \partial_k B_{\alpha,\beta}(k) =
		&- k B_{\alpha,\beta+1}(k) 
		\nonumber\\	
	 - k B_{\alpha-1,\beta}(k) &- 4 K_{\alpha+\beta}(2k)
		\:.
\end{align}
These relations can be used to obtain the behaviour of $B_{\alpha,\beta}(k)$ as $k \to 0$~:
\begin{equation}
	B_{\alpha,\beta}(k) \sim k^{-\xi(\alpha,\beta)}(1 + O(k^{\eta(\alpha,\beta)}))
	\:,
\end{equation}
where
\begin{equation}
	\xi(\alpha,\beta) = \mathrm{max}(|\alpha+\beta|,|\alpha-\beta|)
	\:,
\end{equation}
and
\begin{equation}
	\eta( \alpha , \beta ) = 2\, \mathrm{min}( 1, |\alpha| , |\beta| )
	\:.
\end{equation}

For odd $N$ we must add one line and one column to the matrix $A$ defined for even $N$, as explained above.
Both cases can then be treated on the same footing.

Denote  
\begin{equation}
  \nu = \mu - (N-2n-1)/2 \in ]0,1[
\end{equation}
with $n \in \lbrace 1, \ldots, N \rbrace$. It is easier to compute $\det(A)$ by decomposing the matrix in blocks (for both odd and even $N$)~:
\begin{equation}
	A = \begin{pmatrix}
		A_n & B_{N,n}\\
		-B_{N,n}^T & C_{N,n}
	\end{pmatrix}
	\:,
\end{equation}
where we introduced
\begin{itemize}
	\item $A_n$ of size $n \times n$,
	\item $B_{N,n}$ of size $n \times (N-n)$,
	\item $C_{N,n}$ of size $(N-n) \times (N-n)$.
\end{itemize}
One can then obtain the expansion of $\det(A)$ as $k \to 0$ using the formula for the determinant of a block matrix~:
\begin{equation}
	\det(A) \sim \det(C_{N,n}) \det(A_n + B_{N,n} C_{N,n}^{-1} B_{N,n}^T  )
	\:.
\end{equation}
And finally the behaviour of $\mathcal{C}_{N,1}$ follows from 
$\pf(A) = \sqrt{\det(A)} \sim k^{\sigma}(1 + O(k^\alpha))$, where the exponent controlling the next leading order term is 
\begin{equation}
	\alpha = 2\, \mathrm{min}(\nu, 1-\nu)
	\:.
\end{equation}
This corresponds to the saw behaviour described in the body of the letter (curve of Fig.~\textbf{2} of the letter up to a rescaling, see below).

\subsection{B.3. Case $\beta=4$ (class CII)}

\begin{figure}[!ht]
\centering
\includegraphics[width=0.45\textwidth]{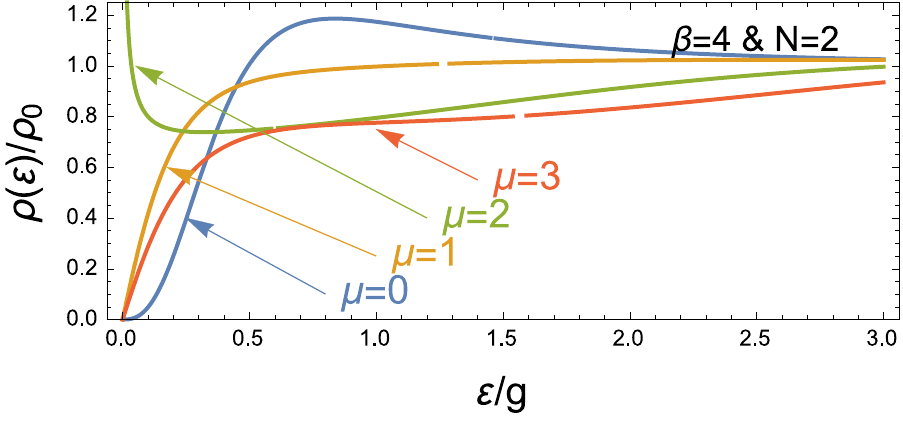}
\caption{\it DoS for the symplectic case for $\Nc=2$ channels. }
\label{fig:DoSBeta4}
\end{figure}

Although we have not explicitly proven that \eqref{eq:SS-SM} is true for the case of matrix with quartionic elements ($\beta=4$), we can compute the multiple integral \eqref{eq:IntegralCN} in this case. 
It can be related to a Pfaffian $\mathcal{C}_{\Nc,4} =  \Nc! \,\pf( A )$ of the $2\Nc\times2\Nc$ matrix 
\begin{equation}
  A_{i,j}
  = (j-i)\int_0^\infty\D z\,\phi(z)\,z^{i+j-3}
  \:.
\end{equation}
Some algebra gives 
\begin{align}
  \label{eq:PfaffianForBeta4}
    \mathcal{C}_{\Nc,4} =& \: \Nc ! \, 2^{\Nc}\, k^{-\Nc\mu}\:
    \\\nonumber
    &\times\mathrm{pf} \big[ (j-i)\,K_{\mu+1+2\Nc-i-j}(k/g) \big]_{1\leq i,\,j\leq2\Nc}
  \:.
\end{align}
The formula obviously give the known result for $\Nc=1$, as it should.
For $\Nc=2$ we check that \eqref{eq:PfaffianForBeta4} leads to the low energy behaviour $\mathcal{N}(\varepsilon)\sim\varepsilon^4\ln(1/\varepsilon)$ and the expected power law (Fig.~\ref{fig:DoSBeta4}).

\subsection{B.4. Low energy DoS -- Summary}

The previous analysis has demonstrated that the exponent $\alpha$ controlling the low energy intergated DoS, $\mathcal{N}(\varepsilon)\sim\varepsilon^\alpha$, presents the saw behaviour as a function of $\mu$.
The dependence in the Dyson index $\beta$ can be simply written under the form
\begin{equation}
  \label{eq:GeneralAlpha}
  \alpha_\beta(\mu) = \frac{\beta}{2}\, \alpha_{2} \left( \frac{2}{\beta} \mu \right)
  \:.
\end{equation}
In the case $\beta=4$ (chiral symplectic case), we have checked the relation with \eqref{eq:PfaffianForBeta4} for $\Nc=2$.


\section{C. Lyapunov spectrum and conductance at a topological phase transition}

Away from the $\Nc$ topological phase transitions, all the Lyapunov exponents are finite. In this case we expect conventional localisation properties, reflected for example in the conductance $\conduc$ characterising the transmission through a slice of random medium (i.e. the random mass is non zero on  $[0,L]$ and vanishes on $]-\infty,0]\cup[L,+\infty[$).
The conductance is expected to present the generic behaviour for 1D or quasi-1D systems~: exponential decay with $L$.
The logarithm of the conductance is expected to be self averaging, thus the distribution of $\ln\conduc$ is expected to be Gaussian, centered on $\mean{\ln\conduc}\sim-2L/\xi_\mathrm{loc}$, where $\xi_\mathrm{loc}$ is the localisation length. 

Precisely at a topological phase transition, when the exponent $\alpha$ vanishes, the model presents atypical localisation properties.
For $\varepsilon=0$, the conductance can be expressed in terms of the variables as $\conduc=\sum_n\cosh^{-2}\zeta_n(L)$ introduced in the paper, where each variable $\zeta_n(x)$ behaves asymptotically as a free Brownian motion with a drift given by the corresponding Lyapunov exponent.
When one Lyapunov exponent vanishes, say $\gamma_n=0$, the conductance is dominated by the contribution of the corresponding variable $\zeta_n$, thus $\argcosh[1/\sqrt{\conduc}]$ is a Gaussian variable with variance $gL$, as in 1D \cite{SteCheFabGog99SM} (this is a new manifestation of superuniversality at the topological phase transition). 
For $L\to\infty$, the moments of the conductance then behave as 
$\mean{\conduc^n}\simeq2^{2n}/(n\sqrt{2\pi gL})$, indicating very large fluctuations, with \textit{non self averaging} $\ln\conduc$ when $L\to\infty$~:  
$\mean{\ln\conduc}\simeq-\sqrt{8gL/\pi}$ and $\mathrm{var}(\ln\conduc)\simeq4(1-2/\pi)gL$
(this was obtained in \cite{BroFurGruMud00SM,BroMudFur01SM} for $\mu=0$ and $N$ odd~; see also \cite{RamTex14SM} for the energy dependence in the case~$N=1$).


\end{document}